# Metal – TiO$_2$ contacts: An electrical characterization study

L. Michalas[*], A. Khiat, S. Stathopoulos, T. Prodromakis

Nanotechnology Research Group, Electronics and Computer Science Department, University of Southampton, Southampton, SO17 1BJ, UK.

[*]Corresponding Author's email:l.michalas@soton.ac.uk

**Abstract:** The electrical properties of thin TiO$_2$ films have recently been extensively exploited towards enabling a variety of metal-oxide electron devices: unipolar/bipolar semiconductor devices and/or memristors. In such efforts, investigations on the role of TiO$_2$ as active material have been the main focus, however, electrode materials are equally important. In this work, we address this need by presenting a systematic quantitative electrical characterization study on the interface characteristics of Metal-TiO$_2$-Metal structures. Our study employs typical contact materials that are used both as top and bottom electrodes in a Metal-TiO$_2$-Metal setting. This allows investigating the effect of Metal-TiO$_2$ contacts as well as holistically studying an electrode's influence to the opposite interface, referred to as top/bottom electrodes coupling. Our methodology comprises the recording of current-voltage (I-V) characteristics from a variety of solid-state prototypes in the temperature range of 300-350 K and by analysing them through appropriate modelling. Clear field and temperature dependent signature plots were also obtained towards shinning more light on the role of each material as top/bottom electrodes in Metal-TiO$_2$-Metal structures. Our results provide a useful database for selecting appropriate electrode materials in Metal-TiO$_2$ devices, offering new insights in metal-oxide electronics applications.

## Introduction

Metal-oxides (MOs) combine a unique ensemble of properties presenting great potential to meet the diverse requirements of modern electronics technologies. Specifically they offer low temperature (<100ºC) manufacturability[1], thus compliance with large-scale uniformity and deployment on alternative materials (paper, flexible); adequate device level mobility[2] that can be further optimized by engineering appropriate gate dielectrics; high transparency due to their wide band gap; capacity for post fabrication tuneable resistance (memristive effects)[3,4] and good chemical stability. The features enabled the use of MOs in a variety of applications ranging from Resistive Random Access Memories (RRAMs)[5,6] and Thin Film Transistors (TFTs)[7,8] to oxide-based photovoltaics[9] and sensors[10], introducing a new era for large area transparent/stretchable electronics[11,12] and neuromorphic systems[13,14]. To this end, a variety of materials have been scrutinised over the recent years with distinct MO and thus properties being exploited according to the application needs, i.e neuromorphic applications favour MO devices that show large memory capacity[4].

$TiO_2$ is without a doubt one of the most celebrated materials through practical implementations of memristors[15], TFTs[16,17], sensors[18] and corollary applications of these. The ability of $TiO_2$ to obtain different microstructures (i.e. amorphous, micro/nano crystalline, rutile, anatase etc.) and thus a plethora of electronic properties that can be determined/controlled by the fabrication and/or biasing conditions augmented its use in practical applications. This was further enhanced by the incorporation of foreign metal elements (doping) in $TiO_2$ thin films that were shown to improve RRAM switching characteristics[19] but also enable both n- and p-type functionalities[20,21]. The latter area is still in its infancy, yet, offers substantial prospects for low-temperature manufacturable electronic systems. Research efforts are thus targeted on addressing outstanding challenges[22], with first proof-of-concept results on the development of bipolar components entirely based on $TiO_2$ p-n homo-junctions[23,24] showcasing the exciting technological potential.

Notwithstanding the importance of $TiO_2$ as active layer, identifying appropriate metal contacts and deciphering their interfacial role is now of paramount importance to a device's electrical behaviour. This role becomes even more important in highly scalable thin-film devices, where device properties strongly rely on interfacial effects. These essentially need to be fully understood for optimising a device's electrical response.

Up to date, such studies are mainly restricted to room temperature semi-quantitative approaches, focusing only in studying electrodes formed atop of $TiO_2$ films. More specifically, it was reported that recording the rectification ratio measured at read out-voltage of ± 1V for a variety of top electrodes (deposited by RF magnetron sputtering on Metal-$TiO_2$-(Ohmic-Pt) configuration) can reveal the role of a metal's electronegativity on the formation of the interface barrier[25]. In the same study, barriers evaluated by the forward diode current, suggested partial Fermi level pinning. In addition, the role of the top electrode (evaporated through shadow mask) on a Metal-$TiO_2$-(Ohmic-Pt) memory cell, evaluated with a read-out voltage of ± 1V, was correlated to the formation free energy for top metal electrode oxide[26]. Moreover, the symmetry/asymmetry of Metal-$TiO_2$-(Ohmic-Al) current-voltage characteristics attributed also to the top metal fabrication details such as the thermal annealing[27]. More recently, the source and drain electrodes of $TiO_2$ based TFTs were found to play a major role on performance parameters such as the ON/OFF ratio and the field effect mobility[17]. All the above assessments suggest that this is not a conventional Metal-Semiconductor contact where the metal work function dominates the process and therefore additional studies are still required for an in-depth investigation.

In this paper, we present a detailed quantitative electrical characterization study of Metal-TiO$_2$ interface characteristics. The work involves the commonly utilized contact metals acting both as top and bottom contacts in Metal-TiO$_2$-Metal configuration, whilst it also reports the influence of the top electrode (TE) material to the bottom electrode (BE) interface. This is performed by recording the current-voltage (I-V) characteristics at different temperatures and through appropriate modelling and analysis that includes field and temperature dependent signature plots. Considering also the potential applications, the design and fabrication of all prototyped samples, on which this study was based, remained aligned to standard low temperature microelectronic processes, while the electrodes have been patterned through optical lithography and deposited via electron-beam evaporation.

**Results**

**Device Modelling and Parameters Extraction:** The studied Metal-TiO$_2$-Metal devices are presented in Fig. 1(a). In order to obtain quantitative results the tested devices have been modelled by the equivalent circuit presented in Fig. 1 (b). This consists of two series resistances accounting for the top ($R_{TE}$) and the bottom ($R_{BE}$) access electrodes and the device under test (DUT). As TiO$_2$ exhibits an intrinsic n-type character, the DUT can be considered as two inverse polarized Schottky diodes, both emulating the TE and BE contacts, connected through a resistance that corresponds to the TiO$_2$ core. This equivalent circuit was further used for calculating the effective applied bias ($V_{eff}$)/electric field across the DUT, with respect to the bias applied through the voltage source (V) and as a function of the measured current (I), summarised by Eq.1

$$V_{eff} = V - I(R_{TE} + R_{BE}) \qquad (1)$$

In the absence of an interface barrier, the Schottky diodes (TE/BE) can be considered as short-circuit. In this case the I-V characteristic should be symmetric with respect to the applied bias polarity and the dominant transport mechanism is purely determined by the properties of the active layer film. For all other cases, where an interface barrier is formed, any positive bias applied to the TE, will result in a forward biased TE/TiO$_2$ and reversed biased TiO$_2$/BE contact. In this case, the current flow will be determined by the reversed biased BE, given that this constitutes the most resistive element in this series configuration. The situation is the opposite when a negative bias is used at the TE; the TE determining the current flow.

The interface-controlled transport is typically identified by an asymmetric I-V characteristic with respect to the applied bias polarity. The transport is then dominated by either tunnelling through the barrier formed at the interface or by thermionic emission over it. Tunnelling is thus expected in cases where carriers do not possess sufficient energy to overcome the potential barrier. The two mechanisms can be distinguished via the temperature dependence of the I-V[28].

Tunnelling currents obey the following Eq. 2:

$$I \propto V_{eff}^2 exp\left(-\frac{b}{V_{eff}}\right) \qquad (2)$$

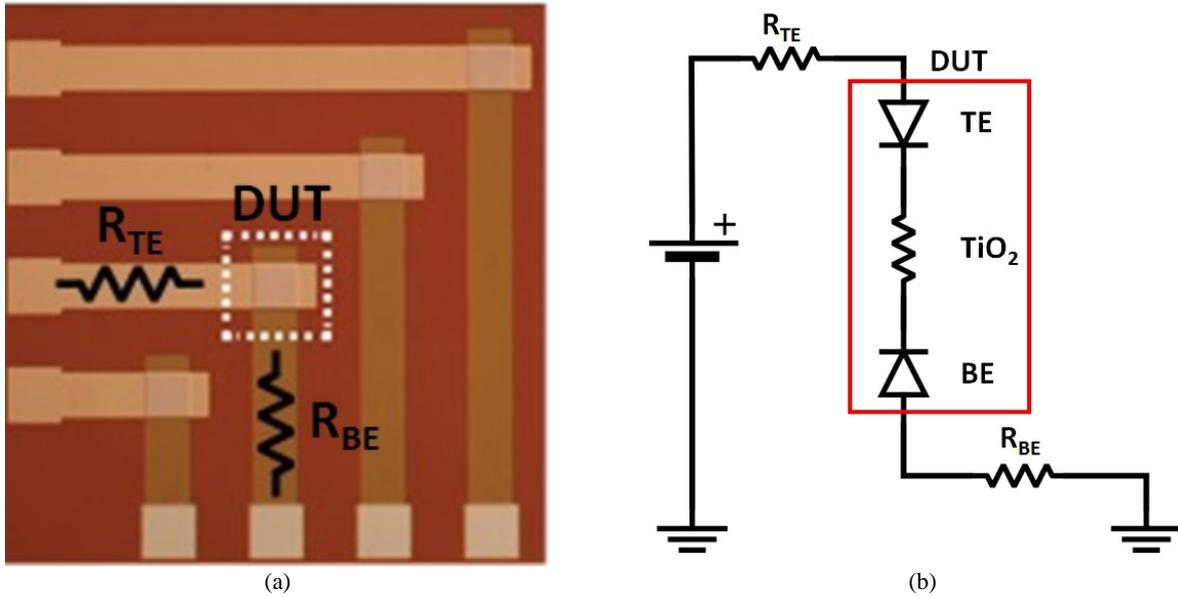

(a)                          (b)

Figure 1: The studied Metal-TiO$_2$-Metal devices (a). The equivalent circuit utilized to model the devices and the electrodes for our quantitative analysis.

with "b" being just a constant. Consequently, an asymmetric, temperature independent I-V, supporting also a linear relation on a ln(I/V$_{eff}^2$) vs 1/V$_{eff}$ signature plot serves as a strong indication of tunnelling dominated transport.

On the contrary thermionic emission over the interface barrier is the most commonly thermally activated mechanism for interface controlled transport and can be described by Eq. 3:

$$I = AT^2 exp\left(-\frac{\Phi_{B0} - a\sqrt{V_{eff}}}{KT}\right) \quad (3)$$

where K is the Boltzmann constant, T is the absolute temperature, $\Phi_{B0}$ is the zero bias potential barrier height at Metal/MO interface and A=(Area x A*), with A* being the Richardson constant and "α" the barrier lowering factor. A straight line on a ln(I/T$^2$) vs 1000/T plot for any applied electric field serves as strong evidence of thermionic emission. The slope of this line corresponds to the apparent interface barrier ($\Phi_{App}$) and should decrease by increasing the applied electric field/bias as:

$$\Phi_{App} = \Phi_{B0} - \alpha\sqrt{V_{eff}} \quad (4)$$

If this $\Phi_{App}$ vs V$_{eff}^{1/2}$ signature plot is also supported by the extracted results, then the intercept extrapolated to V$_{eff}$=0 Volts, provides an experimental determination of $\Phi_{B0,}$ whilst the slope corresponds to "α". By following the aforementioned methodology we can obtain clear signature plots that allow deciphering the transport mechanism and provide quantitative measures of interfaces in Metal-TiO$_2$-Metal structures.

**The Bottom Electrode:** It is important to note that it is possible for the bottom electrode-TiO$_2$ interface to show a distinct behaviour to an identical interface lying at the top of a device, due to processing induced effects. The up to date published works have typically utilized ohmic junctions for the bottom electrode, solely targeting the influence on the top interface characteristics and therefore assuming similar signatures for both interfaces due to symmetry. Nonetheless, we recently demonstrated that even symmetric structures may overall render an asymmetric performance[29].

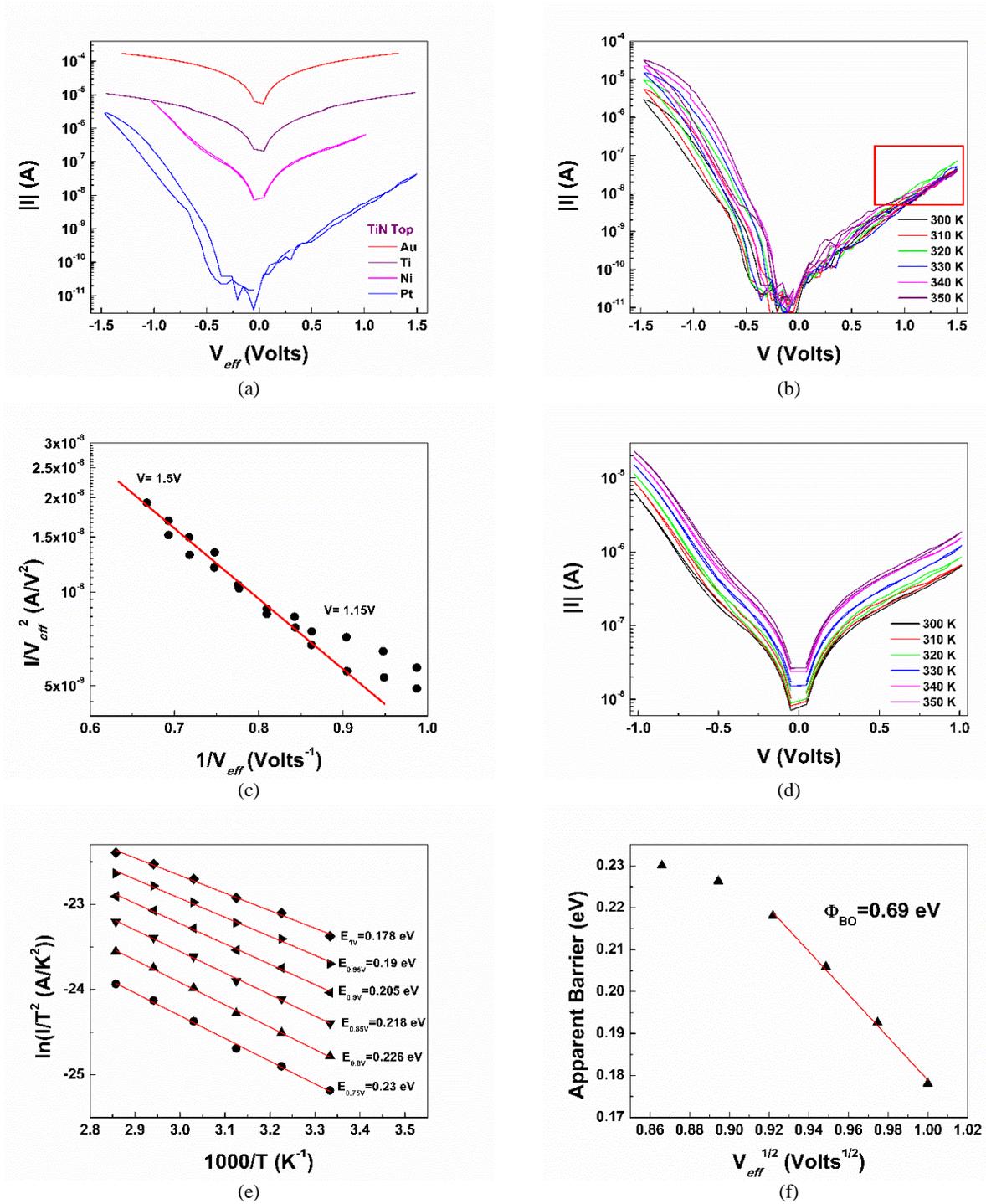

Figure 2: Devices having identical TiN TE and different BE were assessed (a). Tunnelling identified in case of Pt BE signature plots (b),(c) and Schottky emission when Ni is the BE (e), (f).

The I-V characteristics of structures having Au, Ti, Ni and Pt as BE are presented in Fig. 2(a). In all cases the TE is TiN deposited by sputtering. By observation to the acquired symmetric I-Vs, it follows that no interface barrier is formed in the cases where Au and Ti were utilized as bottom electrodes, denoting that the overall conductivity is controlled by the $TiO_2$ thin-film. Besides, the two structures exhibit different resistance measured at read-out voltage of 1V (Table 1), suggesting that the BE material plays a major role on the deposited film conductivity. In cases where Pt and Ni were utilized as BE, strongly asymmetric I-V characteristics where obtained, denoting an interface controlled transport. In these cases, the I-V characteristics have been further assessed in the temperature range of 300 K to 350 K in order to obtain the corresponding signature plots. The applied biases have been limited up to 1.5 V and 1 V respectively to avoid the devices undergoing either a soft or irreversible (hard) breakdown. For the sample with Pt BE (Fig. 2(b)), the part of I-V plot corresponding to the BE (positive biases) presents a temperature independent behaviour (highlighted in the red square). This along with the signature plot presented in Fig. 2(c), supports that tunnelling is the major mechanism responsible for the conductivity, indicating the presence of a very high barrier at this interface that is however immeasurable via Eq. 2.

On the contrary for the device formed atop of Ni BE (Fig. 2(d)), clear signature plots have been obtained that confirm thermionic emission over the interface barrier as the dominant transport mechanism. The data analysis allowed by the clear signature plots, led to the estimation of the zero bias potential barrier at the interface as $\Phi_{B0} = 0.69$ eV. The results for the BE study are summarized in table 1 along with the metal work function and electronegativity as those parameters may determine the interface barrier. We note however that no correlation can be obtained for the BE.

| BE material | $W_f$ (eV) | $X_m$ | Interface Barrier | Additional Information |
|---|---|---|---|---|
| **Au** | 5.1 | 2.54 | No Barrier | R (at 1V) = 8.5 KΩ |
| **Ti** | 4.1 | 1.54 | No Barrier | R (at 1V) = 150 KΩ |
| **Ni** | 5.15 | 1.91 | 0.69 eV | α = 0.511 eV$^{1/2}$ |
| **Pt** | 5.65 | 2.28 | Very high | Tunnelling |

Table: 1: Quantitative results extracted from the experimental data regarding the BE materials

**The Top Electrode:** The TE/TiO$_2$ contacts have been assessed to date mainly through evaluating the DUT's resistance at a specific read-out voltage, typically of 1V. Here we are aiming to present an approach focusing more on quantitative extraction of parameters with physical meaning i.e. the zero bias potential barrier at the interface, $\Phi_{B0}$. For this, several configurations having different TE have been fabricated on the same wafer, with the TiO$_2$ active layer films deposited on top of Au BE during the same process step. Therefore, with the exception to the TE interface, identical characteristics are expected. On the contrary, the results depicted in Fig. 3 provide evidence that the TE deposition plays indeed a major role on the device characteristics.

Ti and Al TE, result in no barrier formation (Fig. 3(a)), while an interface barrier is formed by Au, Pt and Ni (Fig. 3(b)). For these materials, detailed temperature characterization was performed and clear signature plots were extracted by analysing the corresponding I-V curves (Fig. 4). This led to the calculation of several interface quantities (Fig. 5(a)) that are summarized along with resistance of the "ohmic" contacts in Table 2. Considering these experimental results, several interesting features can be noted. Starting from Al and Ti that do not form any interface barrier, it was found that they both affect the conductivity of the structure resulting in a different resistance measured at 1V.

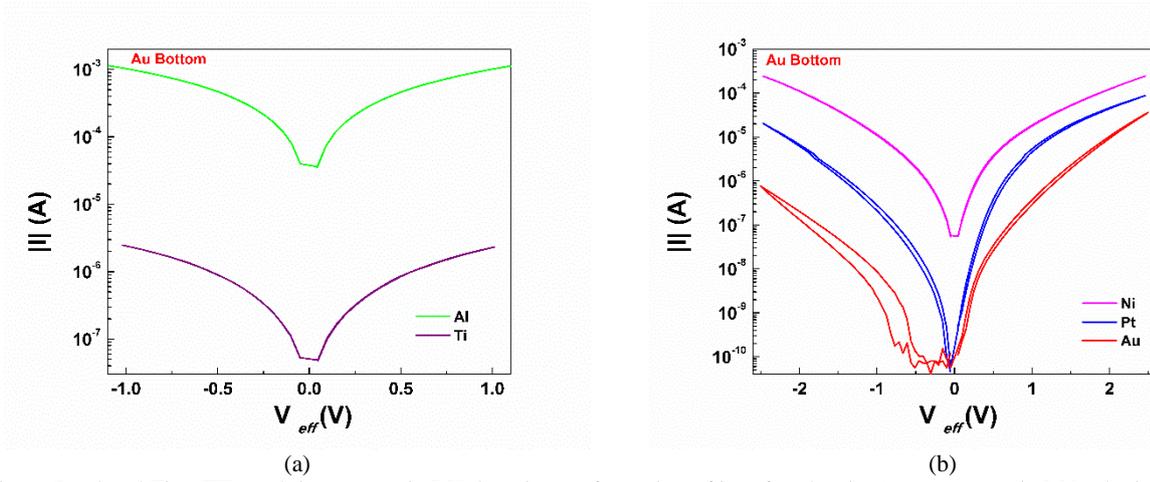

Figure 3: Al and Ti as TE result in symmetric I-V denoting no formation of interface barrier (a). Asymmetric I-Vs obtained in cases of Au, Pt and Ni suggesting interface controlled conduction mechanism (b).

| TE material | $W_f$ (eV) | $X_m$ | Interface Barrier at TE | Additional Information |
|---|---|---|---|---|
| **Al** | 4.28 | 1.61 | No Barrier | R (at 1V) = 1 KΩ |
| **Ti** | 4.1 | 1.54 | No Barrier | R (at 1V) = 430 KΩ |
| **Pt** | 5.65 | 2.28 | 0.61 eV | α = 0.265 eV$^{1/2}$ |
| **Ni** | 5.15 | 1.91 | 0.50 eV | α = 0.208 eV$^{1/2}$ |
| **Au** | 5.1 | 2.54 | 0.70 eV | α = 0.203 eV$^{1/2}$ |

Table 2: Quantitative results extracted from the experimental data regarding the TE materials

This behaviour maybe supported by the mechanism proposed by in Ref. 26 and the Ellingham diagram, where the argument is that electrode metals interact with TiO$_2$ generating oxygen vacancies. Oxygen vacancies act as n-type dopants, providing free electrons in the MO conduction band. Al should induce more oxygen vacancies with respect to Ti and thus results in a more conductive behaviour. Indeed, in cases of "ohmic" contacts and when the transport in the film is dominated by band conduction due to free electrons, the average conductivity is proportional to the free carriers concentration (i.e. σ = qnμ, for n-type material, where n denotes the free electron density, q the electronic charge and μ the band mobility) and thus to the amount of the oxygen vacancies induced by the electrodes. However, this might not be straightforward in cases where the film conductivity is not dominated by the free carriers. In such cases, that also hold for wide band gap materials (which is the expected native condition for TiO$_2$), additional mechanisms should be considered, such as hopping or Frenkel-Poole, depending on the band-gap states density and energy distribution, the operation temperature and the applied field intensity.

The situation becomes more complicated in the case of Ni, Au Pt, where the conductivity is controlled by thermionic emission over the interface barriers, as confirmed by the signature plots. The presence of oxygen vacancies may affect the depletion layer width and thus indirectly the interface barrier. However, the barrier height is also affected by the fermi level position, which is mainly determined by the trap charge at the interface. Moreover, according to Eq. 3 the current conduction depends exponentially on the apparent barrier, including the major role of its lowering due to the applied bias, determined by the interface controlled factor "α". Thus in the present study, the non-ohmic contacts are assessed through the estimation of the zero bias potential barrier, $\Phi_{B0}$, extracted in Fig. 5 (a) through Eq. 4, providing straightforward quantitative information on the interface characteristics.

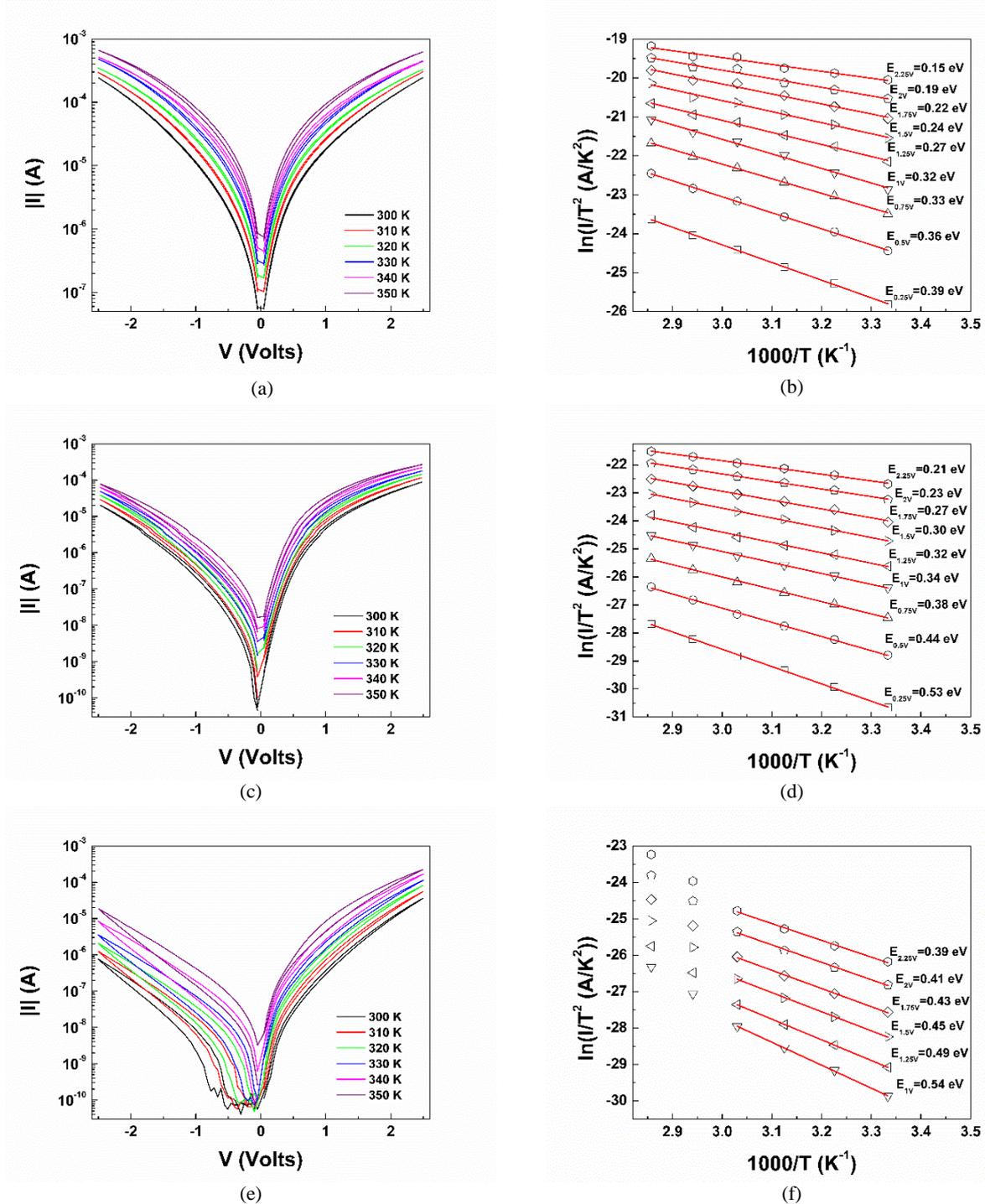

Figure 4: I-V curves vs temperature for Ni (a), Pt (b) and Au (c), TE and the corresponding signature plots (b), (d) and (f) respectively, confirming Schottky emission as the dominant transport mechanism.

Considering the interface barrier, this should be typically proportional to the metal work function. In cases, however, where large amount of defect states are present at the interface, these could significantly affect this dependence. Pinned Fermi position resulting in barrier heights independent of the metal work function are commonly obtained, as for example in the case of III-V compounds. Nonetheless, this does not apply to our work as while the barrier is formed it is not proportional to the metal work function; significant variations exist for distinct materials. The correlation of the formed barrier to the electronegativity has been reported for ionic semiconductors[28], where the strength of the correlation is typically assessed by the index S:

$$S = \frac{d\Phi_{B0}}{dX_m} \quad (5)$$

A strong correlation to electronegativity is considered as the S index approaches 1. For example, Silicon results in S=0.05 while $SiO_2$ to S=1[28]. A clear linear relation is supported by our results depicted in Fig. 5 (b). The calculated index is S= 0.32. Similar results extracted in a completely different way are presented in Ref. 25 that resulted to S= 0.55 including materials, such as Al and Ti, which in our case did not form a barrier. This is referred to as partial Fermi pinning that showcases the important role of interface states.

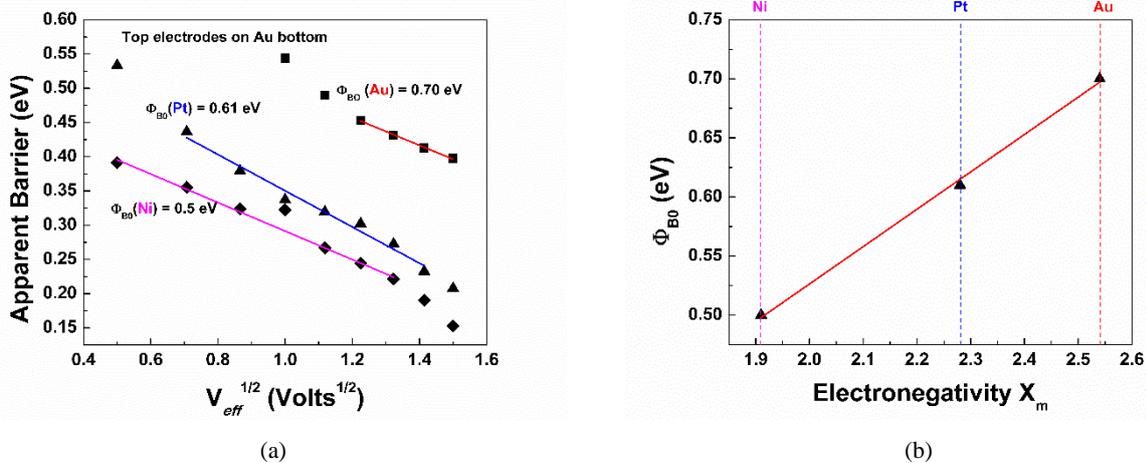

(a) (b)

Figure 5: Experimental verification for the conduction mechanism and extraction of the zero bias potential barrier at the top interface (a). The barrier is found to increase linearly with the electronegativity of the TE material (b).

It is worth noting that while this feature was identified for the TE, it is not the case for the BE (Table 1), highlighting the differences in the formation mechanism for the two interfaces. Furthermore, an issue that has not been pointed out to date, is the obtained coupling between the two interfaces. Throughout this study, we employ this term for highlighting that the deposition of the TE also results in modifications to the BE interfaces (positive biases in the I-V). Considering that the BE and the $TiO_2$ should be identical from the fabrication point of view, we consider this an interesting finding and therefore it is discussed in more detail in the following section.

**Coupling Issues:** The TE deposition is considered to affect the bottom $Au/TiO_2$ contacts (Fig. 3, positive biases) that are nominally identical for all employed prototype devices. In cases of Al and Ti, no barrier is formed and therefore the conductivity is mainly determined by the film properties. In contrast, depositing Ni, Pt or Au as TE, appears to also instigate the formation of a barrier at the bottom interface, offering the opportunity to perform a quantitative study regarding the role of the TE to the BE interface. Therefore, the $BE/TiO_2$ interfaces have been characterized in detail following the previously employed temperature dependent methodology. The obtained results for the $Au/TiO_2$ BE with respect to the TE material are summarized in table 3, while the corresponding signature plots are presented in Fig. 6. A linear relation to the electronegativity was also obtained for the BE interface barriers. In this case, the barriers formed are lower than the TE ones but follow the same trend. The data presented in Fig. 7 support an S= 0.48. This clearly depicts the influence of the TE material throughout the $TiO_2$ all the way to the BE, but also signifies the major role of the interface states.

| TE material | $W_f$ (eV) | $X_m$ | Interface Barrier at BE | Additional Information |
|---|---|---|---|---|
| **Al** | 4.28 | 1.61 | No Barrier | R (at 1V) = 1 KΩ |
| **Ti** | 4.1 | 1.54 | No Barrier | R (at 1V) = 430 KΩ |
| **Pt** | 5.65 | 2.28 | 0.54 eV | α = 0.279 $eV^{1/2}$ |
| **Ni** | 5.15 | 1.91 | 0.37 eV | α = 0.161 $eV^{1/2}$ |
| **Au** | 5.1 | 2.54 | 0.67 eV | α = 0.236 $eV^{1/2}$ |

Table 3: Quantitative results extracted from the experimental data regarding the bottom interface with respect to the TE materials

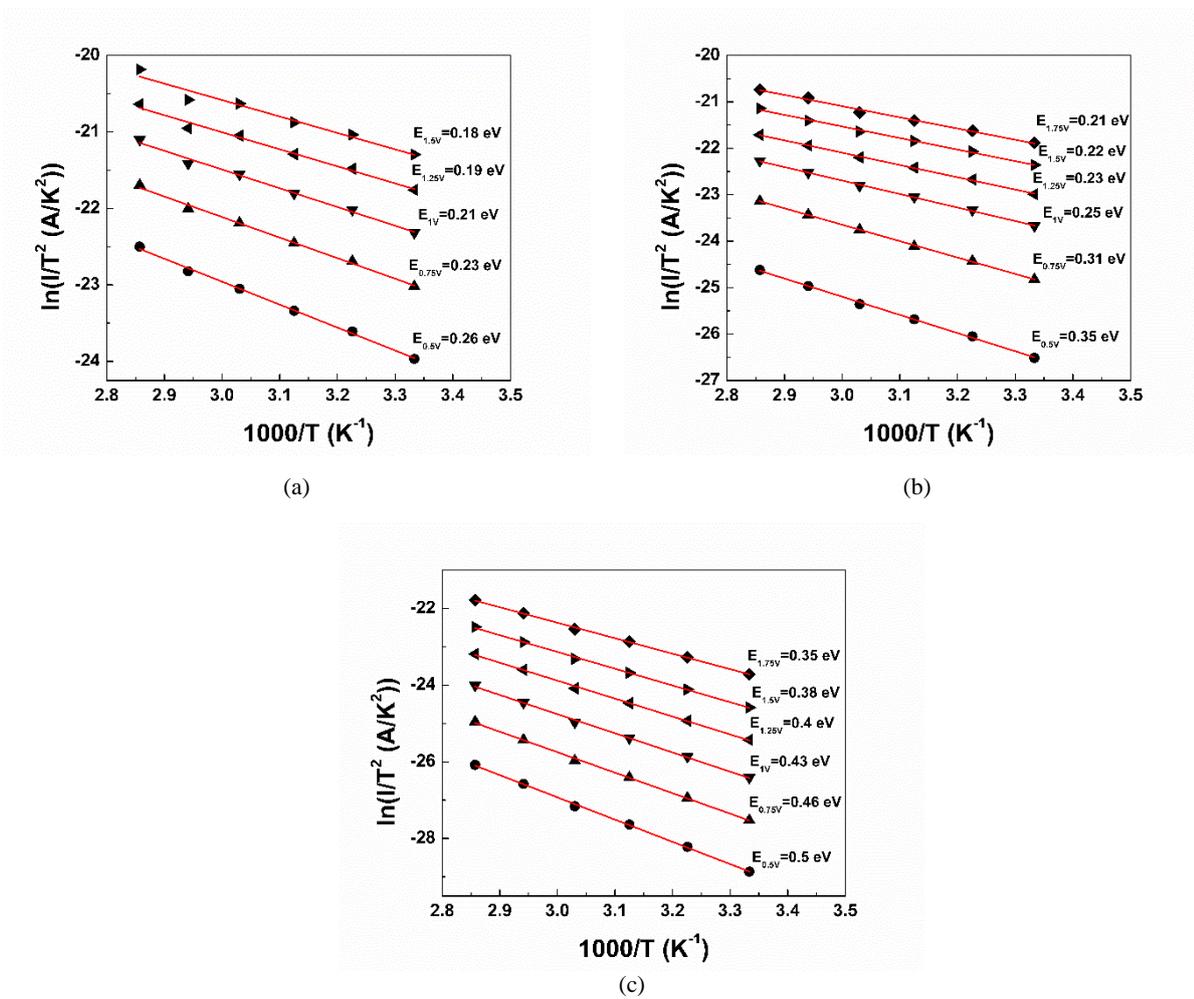

Figure 6: Signature plots (I-Vs at Fig. 4) correspond to the bottom interface, for Ni (a), Pt (b) and Au (c) TE, confirming Schottky emission as the dominant transport mechanism.

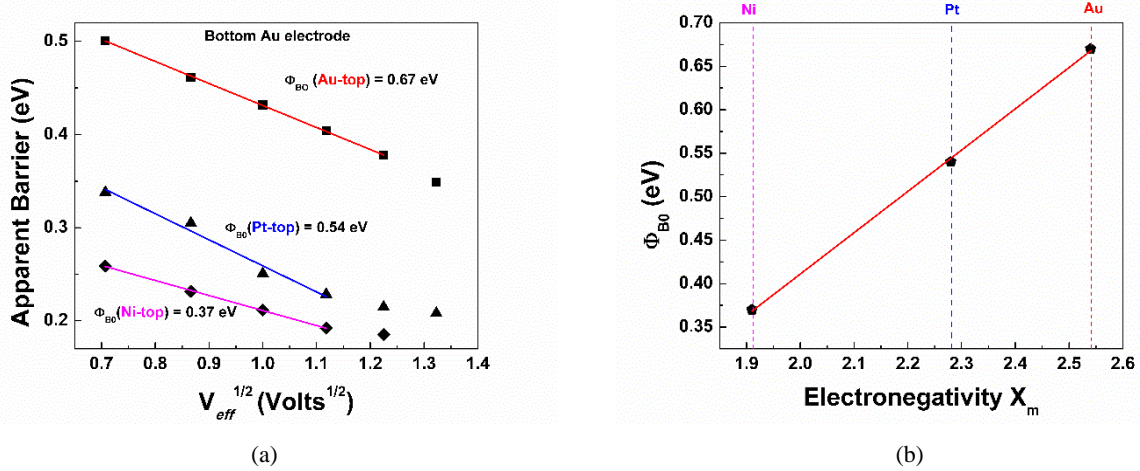

Figure 7: Further experimental verification for the conduction mechanism and extraction of the zero bias potential barrier at the interface for the bottom interface (a). The barrier is found to increase linearly with the electronegativity of the TE material (b).

**Discussion**

This study has provided several new insights on the electrical properties of Metal-$TiO_2$ interfaces. Evidently, these cannot be considered as conventional Schottky contacts, where the conduction is mainly determined by the metal work function. In the case of Au BE, two different regimes were identified for TE. When no interface barrier is formed (Al and Ti) the film conductivity is enhanced for higher formation free energy of the metal oxide, possibly due to an increase of oxygen vacancies, as proposed in Ref. 26. In cases where an interface barrier exists, its height is found to be proportional to the metal electrode electronegativity. The proportionality factor (S index) is far from 1, thus denoting the important role of the interface states. This study also came across an unexpected observation (interface coupling), where the BE interface barrier was found to be determined by the TE material, following accordingly ohmic or electronegativity dependent barrier. This requires that the metal electrodes are selected in pairs, as the TE defines the properties of the bottom interface as well. The opposite trend, i.e. the BE barrier determined by the electronegativity is not however true. It is therefore concluded that different mechanism are responsible for the formation of the interfaces in cases of BE, where the $TiO_2$ film is deposited on the metal, in contrast to the TE interfaces where the metal is deposited atop the $TiO_2$ film. As a result, each material should be carefully considered with respect to its utilization as TE or BE; although in some case like Ti, no interface barrier is formed for either configuration. Furthermore, the device conductivity and the top interface characteristics are also found to be affected by the BE material, supporting our argument that electrode materials should always be selected in pairs.

Considering the growth of $TiO_2$-based electronic devices and the necessity for metal electrodes on top or below their active area, we proceed by sharing some thoughts on the usefulness of our results across the diverse MO applications. In the field of RRAM, where typically the devices are utilized after an electroforming step, Schottky barriers at the TE are considered essential for supporting resistive switching[30]. It was also reported that high barriers at TE support both Unipolar and Bipolar resistive switching, in contrast to lower barriers that typically favour bipolar operation[31]. The proper selection of metal electrodes can be utilized for engineering the electroforming process, in terms of polarity and applied field intensity, and perhaps even developing forming-free RRAM technologies. The electrodes' characteristics are also of major importance to semiconductor devices. For example, when considering TiO$_2$-based bipolar devices it is essential to avoid any rectification stemming from the metal electrodes. Moreover, in the case of TFTs, different configurations may require the Source and Drain electrodes atop or below the deposited MO films defining the TFTs' active area and thus should be accordingly selected. Finally, several applications require the monolithic integration of such technologies. A prominent example is the use of diodes/transistors as selecting elements[32] for RRAM, rendering 1D1R and/or 1T1R configurations. Such cases typically exploit common metal electrodes acting both as BE and TE for different devices and thus should be accordingly considered, bearing also in mind the coupling issues.

## Methods

**Device Fabrication:** The tested devices were fabricated on two separate 6-inch Si wafers having a 200 nm thick SiO$_2$ layer that was grown by dry oxidation. All electrodes and active areas where patterned via standard optical lithography. The first wafer was dedicated for studying the BE influence, with four different areas utilised for depositing 15 nm of Ti, Ni, Pt and Au. These metals were deposited via electron beam evaporation, on top of 5 nm thick Ti adhesion layer. The second wafer enabled the study of the TE and thus Au was utilized as BE. 25 nm of amorphous TiO$_2$ films were deposited by reactive sputtering (Helios XP, Leybold optics) using a Ti target in oxygen plasma environment. 15nm thick TiN was sputtered everywhere across the first wafer, serving as TE. For the second wafer five separate areas were defined and 15nm thick Al, Ti, Ni, Pt and Au films were respectively deposited via electron beam evaporation.

**I-V measurements:** The current vs voltage (I-V) characteristics were obtained on 30 x 30 μm$^2$ devices using our in-house memristor characterization platform ArC ONE$^{TM}$ [33] . The voltage sweeping was carried out in staircase mode towards positive biases, while both positive and negative polarities were always applied to the top electrode (TE) with respect to the bottom electrode (BE) that was continuously kept grounded. All experiments were performed on a Cascade SUMMIT 12000B semi-automatic probe station that incorporates a thermal chuck, whose temperature can be controlled by an ESPEC ETC-200L unit. Measurements were performed in the temperature range of 300 K to 350 K.

## References

1. Yu, X., Marks, T. J. & Facchetti, A. Metal oxides for optoelectronic applications. *Nat. Mater.* **15,** 383–396 (2016).
2. Brox-Nilsen, C., Jidong, J., Yi, L., Peng, B. & Song, A. M. Sputtered ZnO Thin-Film Transistors With Carrier Mobility Over 50 cm$^2$/ Vs. *Electron Devices, IEEE Trans.* **60,** 3424–3429 (2013).
3. Kim, W. *et al.* Multistate Memristive Tantalum Oxide Devices for Ternary Arithmetic. *Sci. Rep.* **6,** 36652 (2016).
4. Stathopoulos, S. *et al.* Multibit memory operation of metal-oxide bi-layer memristors - Supplementary information. *ArXiV* **1704.03313,** 2–8 (2017).
5. Ielmini, D. Resistive switching memories based on metal oxides: mechanisms, reliability and scaling.

*Semicond. Sci. Technol.* **31,** 063002 (2016).

6. Sawa, A. Resistive switching in transition metal oxides. *Mater. Today* **11,** 28–36 (2008).
7. Park, J. S., Maeng, W. J., Kim, H. S. & Park, J. S. Review of recent developments in amorphous oxide semiconductor thin-film transistor devices. *Thin Solid Films* **520,** 1679–1693 (2012).
8. Kamiya, T., Nomura, K. & Hosono, H. Origins of High Mobility and Low Operation Voltage of Amorphous Oxide TFTs: Electronic Structure, Electron Transport, Defects and Doping. *J. Disp. Technol.* **5,** 273–288 (2009).
9. Rühle, S. *et al.* All-oxide photovoltaics. *J. Phys. Chem. Lett.* **3,** 3755–3764 (2012).
10. Fine, G. F., Cavanagh, L. M., Afonja, A. & Binions, R. Metal oxide semi-conductor gas sensors in environmental monitoring. *Sensors* **10,** 5469–5502 (2010).
11. Petti, L. *et al.* Metal oxide semiconductor thin-film transistors for flexible electronics. *Appl. Phys. Rev.* **3,** (2016).
12. Liu, Y., Pharr, M. & Salvatore, G. A. A Lab-on-Skin: A Review of Flexible and Stretchable Electronics for Wearable Health Monitoring. *ACS Nano* acsnano.7b04898 (2017). doi:10.1021/acsnano.7b04898
13. Serb, A. *et al.* Unsupervised learning in probabilistic neural networks with multi-state metal-oxide memristive synapses. *Nat. Commun.* **7,** 12611 (2016).
14. Mehonic, A. & Kenyon, A. J. Emulating the electrical activity of the neuron using a silicon oxide RRAM cell. *Front. Neurosci.* **10,** (2016).
15. Strukov, D. B., Snider, G. S., Stewart, D. R. & Williams, R. S. The missing memristor found. *Nature* **459,** 1154–1154 (2009).
16. Shih, W. S., Young, S. J., Ji, L. W., Water, W. & Shiu, H. W. $TiO_2$-Based Thin Film Transistors with Amorphous and Anatase Channel Layer. *J. Electrochem. Soc.* **158,** H609 (2011).
17. Choi, H., Shin, J. & Shin, C. Impact of Source/Drain Metal Work Function on the Electrical Characteristics of Anatase $TiO_2$ -Based Thin Film Transistors. *ECS J. Solid State Sci. Technol.* **6,** 379–382 (2017).
18. Bai, J. & Zhou, B. Titanium dioxide nanomaterials for sensor applications. *Chem. Rev.* **114,** 10131–10176 (2014).
19. Trapatseli, M. *et al.* Engineering the switching dynamics of $TiO_x$-based RRAM with Al doping. *J. Appl. Phys.* **120,** (2016).
20. Zhao, L., Park, S. G., Magyari-Köpe, B. & Nishi, Y. Dopant selection rules for desired electronic structure and vacancy formation characteristics of $TiO_2$ resistive memory. *Appl. Phys. Lett.* **102,** (2013).
21. Anitha, V. C., Banerjee, A. N. & Joo, S. W. Recent developments in $TiO_2$ as n- and p-type transparent semiconductors: synthesis, modification, properties, and energy-related applications. *J. Mater. Sci.* **50,** 7495–7536 (2015).
22. Wang, Z., Nayak, P. K., Caraveo-Frescas, J. A. & Alshareef, H. N. Recent Developments in p-Type Oxide Semiconductor Materials and Devices. *Adv. Mater.* **28,** 3831–3892 (2016).
23. Hazra, A., Chattopadhyay, P. P. & Bhattacharyya, P. Hybrid Fabrication of Highly Rectifying p - n Homojunction Based on Nanostructured $TiO_2$. *IEEE Electron Device Lett.* **36,** 505–507 (2015).

24. Vasu, K., Sreedhara, M. B., Ghatak, J. & Rao, C. N. R. Atomic Layer Deposition of p-Type Epitaxial Thin Films of Undoped and N-Doped Anatase $TiO_2$. *ACS Appl. Mater. Interfaces* **8,** 7897–7901 (2016).
25. Zhong, N., Shima, H. & Akinaga, H. Rectifying characteristic of Pt/$TiO_x$/metal/Pt controlled by electronegativity. *Appl. Phys. Lett.* **96,** (2010).
26. Yang, J. J. *et al.* Metal/$TiO_2$ interfaces for memristive switches. *Appl. Phys. A Mater. Sci. Process.* **102,** 785–789 (2011).
27. Hossein-Babaei, F., Lajvardi, M. M. & Alaei-Sheini, N. The energy barrier at noble metal/$TiO_2$ junctions. *Appl. Phys. Lett.* **106,** (2015).
28. S.M., Sze & K.K., Ng *Physics of Semiconductor Devices*. (John Wiley & Sons, 2006).
29. Michalas, L. *et al.* Interface asymmetry induced by symmetric electrodes on metal-Al:$TiO_x$-metal structures. *Nanotechnology, IEEE Trans.* - accepted
30. Hernández-Rodríguez, E. *et al.* Effect of electrode type in the resistive switching behaviour of $TiO_2$ thin films. *J. Phys. D. Appl. Phys.* **46,** (2013).
31. Kim, W.-G. & Rhee, S.-W. Effect of the top electrode material on the resistive switching of $TiO_2$ thin film. *Microelectron. Eng.* **87,** 98–103 (2010).
32. Cortese, S., Khiat, A., Carta, D., Light, M. E. & Prodromakis, T. An amorphous titanium dioxide metal insulator metal selector device for resistive random access memory crossbar arrays with tunable voltage margin. *Appl. Phys. Lett.* **108,** (2016).
33. Berdan, R. *et al.* A μ-Controller-Based System for Interfacing Selectorless RRAM Crossbar Arrays. *IEEE Trans. Electron Devices* **62,** 2190–2196 (2015).

**Acknowledgements**

This work was supported by the EPSRC Grant EP/K017829/1.

**Author Contribution**

L.M and T.P planned the experiments, L.M. performed the electrical characterization and analysed the results. A.K and S.S. optimised the processes and fabricated the samples. L.M. and T.P. wrote the manuscript. A.K. and S.S. provided useful suggestions for improving it.

**Data Availability**

All data supporting this study are openly available from the University of Southampton repository at: (to be added in the final form)

**Additional Information**

**Competing Interests:** The authors declare that they have no competing interests